\documentclass[11pt]{article}
\usepackage{hyperref}
\pdfoutput=1

\begin{document}
\title{Patterns Formation in Drying Drops of Blood}
\author{D. Brutin, B. Sobac, B. Loquet and J. Sampol \\
\\\vspace{6pt} IUSTI Laboratory - Université de Provence - Marseilles - FRANCE}
\maketitle
\begin{abstract}
The drying of a drop of human blood exhibits coupled physical mechanisms, such as Marangoni flow, evaporation and wettability. The final stage of a whole blood drop evaporation reveals regular patterns with a good reproducibility for a healthy person. Other experiments on anaemic and hyperlipidemic people were performed, and different patterns were revealed. The flow motion inside the blood drop is observed and analyzed with the use of a digital camera: the influence of the red blood cells (RBCs) motion is revealed at the drop periphery as well as its consequences on the final stage of drying. The mechanisms which lead to the final pattern of the dried blood drops are presented and explained on the basis of fluid mechanics in conjunction with the principles of haematology. The blood drop evaporation process is evidenced to be driven only by Marangoni flow. The same axi-symetric pattern formation is observed, and can be forecast for different blood drop diameters. The evaporation mass flux can be predicted with a good agreement, assuming only the knowledge of the colloids mass concentration.   
\end{abstract}

The movie present the evaporation of a drop of whole blood under room conditions on a glass microscope plate. Further details are available in \cite{brutin10}.

\bibliographystyle{plain}



\end{document}